\documentclass[%
 reprint,
superscriptaddress,
 amsmath,amssymb,
 aps,
]{revtex4-1}

\usepackage{booktabs}
\usepackage{multirow}

\usepackage{subcaption}

\usepackage{bm} 
\usepackage{graphicx}
\usepackage{dcolumn}
\usepackage{bm}
\usepackage{hyperref}

\begin{document}

\preprint{APS/123-QED}

\title{Periodic orbits and gravitational wave radiation in short hair black hole spacetimes for an extreme mass ratio system}
\author{Lai Zhao}

\author{Meirong Tang}

\author{Zhaoyi Xu}%
\email{zyxu@gzu.edu.cn(Corresponding author)}
\affiliation{%
 College of Physics,Guizhou University,Guiyang,550025,China
}%


\begin{abstract}

For a short hair black hole(BH) that bypasses the “no-short-hair” theorem, its hair parameters have a significant impact near the event horizon and may thus carry the internal information of the BH. This paper studies the influence of the hair parameter \(Q_m\) and the structural parameter \(k\) of the short hair BH on the periodic orbits of particles and gravitational wave radiation in the extreme mass ratio(EMR) system.
The research results show that \(Q_m\) has a significant impact on the radius of the orbit, angular momentum and the \(E-L\) space. A larger value of \(k\) weakens this effect and makes it degenerate with the Schwarzschild BH. In addition, an increase in \(Q_m\) or a decrease in \(k\) will lead to a reduction in the period of gravitational waves and a significant enhancement of the signal amplitude.
These results provide new observational means for distinguishing short hair BH from classical BH, and also offer new insights for verifying the no-hair theorem and understanding the physical behaviors near the event horizon.

\begin{description}
\item[Keywords]
No-short-hair theorem; Short hair black hole; Periodic orbit; Gravitational wave
\end{description}
\end{abstract}

\maketitle


\section{\label{sec:level1}Introduction}
General relativity(GR) successfully predicts the existence of BHs and is supported by a wealth of astrophysical observations. For instance, the gravitational wave signals produced by the merger of binary BHs \cite{LIGOScientific:2016emj,LIGOScientific:2016aoc,LIGOScientific:2018mvr,LIGOScientific:2020aai}, as well as the capture of BH shadow images and the observation of the supermassive BH at the center of the Milky Way \cite{EventHorizonTelescope:2019dse,EventHorizonTelescope:2022wkp,Genzel:2010zy}. These discoveries strongly confirm the success of GR. However, within the framework of classical GR, the inevitability of spacetime singularities \cite{Penrose:1964wq} and the problem of information loss caused by the event horizon \cite{Hawking:2005kf} reveal the limitations of GR when dealing with extreme gravitational fields. These limitations prompt researchers to explore possible modifications or develop new gravitational theories, such as loop quantum gravity  (see, e.g.,\cite{Ashtekar:2006rx,Ashtekar:2006wn,Ashtekar:2003hd,Singh:2003au,Wilson-Ewing:2016yan,Lewandowski:2022zce}), string theory (see, e.g., \cite{Witten:1995ex,Horowitz:1996nw,Brandenberger:1988aj}), and theories of the coupling of gravity and matter fields  (see, e.g.,\cite{Bizon:1990sr,Brown:1997jv,Bakopoulos:2023hkh}). These theories provide new perspectives for understanding phenomena in extreme gravitational fields.  However, verifying the rationality of these theories requires a suitable experimental platform, and the extreme gravitational field environment around BHs is an ideal natural experimental site for testing these new theories. The extreme gravitational field near BHs can not only generate gravitational waves and BH shadows but also be accompanied by many complex astronomical phenomena, including gravitational lensing effects  (see, e.g.,\cite{Zhao:2024hep,Zhao:2024elr,Tsukamoto:2016jzh}, photon rings \cite{Yang:2024utv}), and bound orbits. These phenomena provide extremely rich information for studying the spacetime structure near BHs and provide important opportunities for testing gravitational theories in strong gravitational fields.The motion of massless photons along null geodesics determines the shape of the black hole shadow and is closely related to the gravitational lensing effect. The timelike geodesics of massive particles describe the orbital evolution of objects in strong gravitational fields, and are particularly important for characterizing the gravitational wave radiation in binary systems and EMR systems. Therefore, in-depth study of geodesics not only helps to test GR and its modified theories but also helps to further understand the behavior near BHs, especially near the event horizon.

As is well known, the bound orbits of timelike geodesic particles can be divided into precessing orbits and periodic orbits. The early precession of Mercury's perihelion \cite{Will1982TheoryAE} provided one of the early evidences for the success of GR. Its precession results from the weak effect of the solar gravitational field. In strong gravitational fields, the precession phenomenon of stars orbiting supermassive BHs has been widely studied, especially the stellar orbits around Sgr A*. The references are  (see, e.g.,\cite{Hees:2017aal,Will:2007pp,Iorio:2011zi,Grould:2017bsw,GRAVITY:2019tuf,GRAVITY:2020gka}). Among them, the observation of the precession of the star S2 around the massive Sgr A* BH \cite{GRAVITY:2020gka} has been widely applied in testing other theories and constraining BH parameters \cite{Hees:2017aal,QiQi:2024dwc,DellaMonica:2023dcw,Li:2024tld}.In contrast, periodic orbits are more complex. The motion characteristics of these orbits often exhibit zoom-whirl behavior. Therefore, they can encode the orbital information around BHs through the orbits \cite{Levin:2008mq}. These periodic orbits are often classified by the orbital classification method in reference \cite{Levin:2008mq}. In their literature, it is pointed out that periodic orbits can often use rational or irrational numbers $q$ to encode their orbits. Here, $q$ is related to the angular frequency and the radial frequency. When q is an irrational number, the bound orbit at this time shows non-periodic orbits and precession behavior. When $q$ is a rational number, the orbit at this time shows periodic orbits. The orbit corresponding to an irrational number can always be well approximated by a periodic orbit \cite{Levin:2008mq}. This also leads people to generally believe that periodic orbits can provide more information about the strong field region than precession. Therefore, a large number of studies on periodic orbits using classification methods have emerged. For example, studies on periodic orbits around Schwarzschild BH, Reissner-Nordström BH, and Kerr BH \cite{Lim:2024mkb,Levin:2008mq,Misra:2010pu,Levin:2009sk,Grossman:2011ps,Grossman:2011im}. In addition, studies on periodic orbits in other spacetime backgrounds are also gradually enriched  (see, e.g.,\cite{Lin:2023rmo,Deng:2020yfm,Zhou:2020zys,Wei:2019zdf,Mummery:2022ana,Babar:2017gsg}). 

For stars (or compact objects) orbiting supermassive BHs, they usually form EMR systems. The smaller celestial body (test particle) gradually loses energy and angular momentum in the strong gravitational field of the BH and spirals closer to the BH. This energy is released through gravitational wave radiation. The gravitational waves radiated by extreme mass ratio systems have once become one of the detection targets of detectors such as LISA, Taiji, and Tianqin due to their high event rate \cite{LISA:2017pwj,Danzmann:1997hm,Gong:2021gvw,Zi:2021pdp}. Through precise observation of the gravitational wave signals radiated by these systems, the properties of BHs and potential cosmological information can be further revealed \cite{LISA:2022kgy}.
In EMR systems, as small celestial bodies spiral down, different periodic orbits act as transition orbits. The study of these transition orbits is crucial in the calculation of gravitational wave radiation \cite{Glampedakis:2002ya}. As described in references \cite{Grossman:2011im,Levin:2008mq}, the analysis of periodic orbits can quickly calculate the gravitational wave waveforms of adiabatic EMR systems, which has also been widely used in recent references \cite{Li:2024tld,Yang:2024lmj,Tu:2023xab}. Therefore, further study of periodic orbits and the use of adiabatic approximation within a single period to construct gravitational wave detection are expected to provide a further understanding of the extreme environment around BHs.


Furthermore, for a short hair BH that violates the no-hair theorem and circumvents the “no-short-hair” theorem \cite{Brown:1997jv}, its main feature lies in generating arbitrarily short "hairs" near the event horizon by taking into account the coupling between anisotropic matter and the Einstein gravitational field. These hairs are usually accompanied by quantum effects and may carry important information about the internal structure and evolution of BHs \cite{Brown:1997jv}. Especially near the event horizon, quantum fluctuations and gravitational perturbations may significantly affect the local spacetime structure. Therefore, studying timelike geodesics, periodic orbits and their gravitational wave radiation characteristics in this spacetime background will help to gain a deeper understanding of the properties of the short hair BH.

This paper mainly focuses on comparing the dynamical behaviors of classical Schwarzschild BHs and short hair BH in the EMR system. By thoroughly analyzing the differences between these two types of BH in terms of timelike geodesics, periodic orbits and the characteristics of gravitational wave radiation, it aims to reveal the unique manifestations of short hair BHs under extreme gravitational fields. Such a comparison not only helps to evaluate the applicability of the short hair BH model in explaining observational phenomena but also provides a new theoretical basis for testing the modified theories of general relativity.
Specifically, we explore the characteristics of gravitational wave signals of the short hair BH in the EMR system and conduct a comparative analysis with the signals of the classical Schwarzschild BH to investigate the influence of hair parameter on these signals. Through this comparative study, it is expected to provide some theoretical references for future astronomical observations and promote a deeper understanding of the extreme gravitational field environment around BHs. Such gravitational wave signals not only offer a new observational window for BH dynamics but may also reveal the internal information of BHs and their evolution processes, thereby providing a new perspective for resolving the BH information paradox and advancing the development of the quantum gravity theory.

The structure of this paper is arranged as follows. In Section \ref{sec:level2}, we introduce the background and spacetime structure of short hair BHs, especially the detailed information of timelike geodesics. In Section \ref{sec:level3}, we derive how the angular momentum and orbital radius of bound orbits are affected by the hair parameter $Q_m$ and parameter $k$, and analyze the energy-angular momentum ($E-L$) space allowed by bound orbits. In Section \ref{sec:level4}, we plot several typical periodic trajectories and discuss the influence of the hair parameter $Q_m$ and parameter $k$ on these trajectories. In Section \ref{sec:5}, based on the kludge waveform model developed in reference \cite{Babak:2006uv}, we show the gravitational wave signal of a single period. In the last section, we summarize the main results of this paper. Unless otherwise specified, this paper uses natural units, that is, $G = c = 1$.

\section{\label{sec:level2}Short Hair Black Hole and Timelike Geodesics}
The no-hair theorem states that a BH can be described by just three parameters: mass \( M \), angular momentum \( J \), and electric charge \( Q \) \cite{Ruffini:1971bza,Hawking:1971vc,Israel:1967wq}.
However, when the exterior of a BH is filled with various matter fields, the no-hair theorem will be violated. The first BH solution that violates the no-hair theorem was calculated by Piotr Bizon. He obtained a “charged” BH solution through numerical analysis methods \cite{Bizon:1990sr}.
The emergence of this BH solution that violates the no-hair theorem has attracted extensive attention in the academic community and prompted researchers to actively explore more BH solutions with hair. In this regard, Ovalle and Contreras et al. successfully constructed static spherically symmetric and axially symmetric hairy BH solutions through the gravitational decoupling method \cite{Ovalle:2020kpd,Contreras:2021yxe}. Further research has obtained more BH solutions with hair by introducing different types of matter fields. These achievements are reported in relevant literatures respectively  (see, e.g., \cite{Carames:2023pde,Bakopoulos:2023hkh,Zhang:2022csi,Brihaye:2015qtu,Lavrelashvili:1992ia,Hod:2011aa,Tahamtan:2020mbb,Creminelli:2020lxn,Herdeiro:2020xmb,Clough:2019jpm,Brown:1997jv}).
Among them, the authors of reference \cite{Brown:1997jv} obtained a class of static spherically symmetric BH solutions with short hair by considering the coupling of Einstein gravity and anisotropic matter fields. The hair of this kind of solution only has a significant influence near the event horizon and does not violate the "no-short-hair" theorem. The detailed reasons will be discussed later.

In the context of anisotropic matter, J. David Brown and Viqar Husain constructed static spherically symmetric black hole solutions with “short hair” by coupling Einstein gravity with anisotropic matter fields. The spacetime metric can be written as \cite{Brown:1997jv}
\begin{equation}
ds^2=-f(r)dt^2+\frac{1}{f(r)}dr^2+r^2(d\theta^2+sin^2\theta d \phi^2),
\label{1}
\end{equation}
where
\begin{equation}
f\left(r\right)=1-\frac{2M}{r}+\frac{Q_m^{2k}}{r^{2k}}.
\label{2}
\end{equation}
Here, $Q_m$ is a parameter that describes the strength of the BH's hair. When the hairy parameter disappears ($Q_m = 0$), the short hair BH degenerates into a Schwarzschild BH. Different $k$ values correspond to different BH spacetime structures. For this short hair BH, its energy density is
\begin{equation}
\rho=\frac{Q_m^{2k}\left(2k-1\right)}{8\pi r^{2k+2}}.
\label{3}
\end{equation}
According to the original literature \cite{Brown:1997jv}, this kind of short hair BH satisfies the weak energy condition, that is, $\rho > 0$. This means that the parameter $k$ must be greater than 1/2. In addition, this short hair BH does not violate the "no-short-hair" theorem.

The "no-short-hair" theorem states that if a static spherically symmetric BH carries hair, then the influence of the hair should not be limited only to the vicinity of the event horizon, but should at least extend beyond the photon sphere radius or beyond 3/2 times the event horizon radius \cite{Nunez:1996xv}.
Of course, there are also scholars who generalize the “no-short-hair” theorem to non-spherically symmetric cases. That is to say, under static and axisymmetric conditions or within a broader framework of gravitational theory, the influence of the hairy effect should extend beyond the photon sphere, and this conclusion still holds true \cite{Acharya:2024kvv,Ghosh:2023kge}.
However, the prerequisite of the "no-short-hair" theorem is that matter must satisfy the weak energy condition, the energy density $\rho$ must decrease faster than $r^{-4}$, and the trace of the energy-momentum tensor $T = T^{\mu}_{\mu} < 0$ \cite{Nunez:1996xv}. Regarding the condition of the trace of the energy-momentum tensor, some people have extended the condition of the trace of the energy-momentum tensor to be non-negative. However, their validity lies in regular matter \cite{Peng:2020hkz}, which does not violate the situation of anisotropic matter discussed in this article. For the short hair BH discussed in this article, the trace of its energy-momentum tensor is $T = 2\rho(k - 1)$. Obviously, when the parameter $k\geq1$, this kind of short hair BH does not satisfy the prerequisite of the "no-short-hair" theorem (here the trace $T\geq0$), so it is not in contradiction with the "no-short-hair" theorem. This shows that the "no-short-hair" theorem does not apply to the case of anisotropic matter.

In the case where the short hair BH is not in contradiction with the "no-short-hair" theorem, the value of the parameter $k$ at this time should be $k = 1, 1.5, 2,\ldots$. When the parameter $k = 1$, the metric (\ref{1}) at this time becomes the Reissner-Nordström BH ($f(r)=1-\frac{2M}{r}+\frac{Q^2}{r^2}$). When the parameter $k = 1.5$, the metric (\ref{1}) at this time is a short hair BH. When the parameter $k = 2$, the metric (\ref{1}) at this time can be written as
\begin{equation}
f\left(r\right)=1-\frac{2M}{r}+\frac{Q_m^4}{r^4}.
\label{4}
\end{equation}
Such a metric also represents a BH carrying short hair. It is worth mentioning that such a short hair BH has the same metric form as the quantum-corrected BH derived by some scholars under quantum gravity theory \cite{Lewandowski:2022zce}, that is, $f(r)=1-\frac{2M}{r}+\frac{\alpha M^2}{r^4}$.
The characteristic of this quantum-corrected BH is that it successfully avoids the appearance of spacetime singularities. Instead, when the density of matter collapse reaches the Planck scale, the collapse of matter will no longer continue but will rebound and enter an expansion period.
For the short hair BH in this paper, if the value of the parameter $k$ is 2 and the short hair parameter $Q_m^4$ is analogized to $\alpha M^2$, then the metric of the short hair BH at this time has the same form as that of the quantum-corrected BH. This is not a coincidence but is easily explained. This is because the influence of the hair parameter of the short hair BH is mainly distributed near the event horizon. In the extreme region near the event horizon, quantum effects will become very significant.
Therefore, in this paper, we will focus on the short hair BH solutions with $k$ values of $1$, $1.5$, and $2$ (especially discuss the characteristics of $k = 1.5$), and explore how the hair parameters of these BHs change the orbital dynamics of the EMR binary system and the characteristics of its gravitational wave radiation by influencing the spacetime geometry. Through in-depth discussions of these phenomena, we hope to further reveal the physical information inside the BH event horizon, especially the short hair parameters with significant influence near the event horizon and the quantum effects that may carry internal information. These studies not only contribute to understanding the physical properties of short hair BHs but also may provide a new perspective for the exploration inside the event horizon.

\begin{figure*}[]
\includegraphics[width=1 \textwidth]{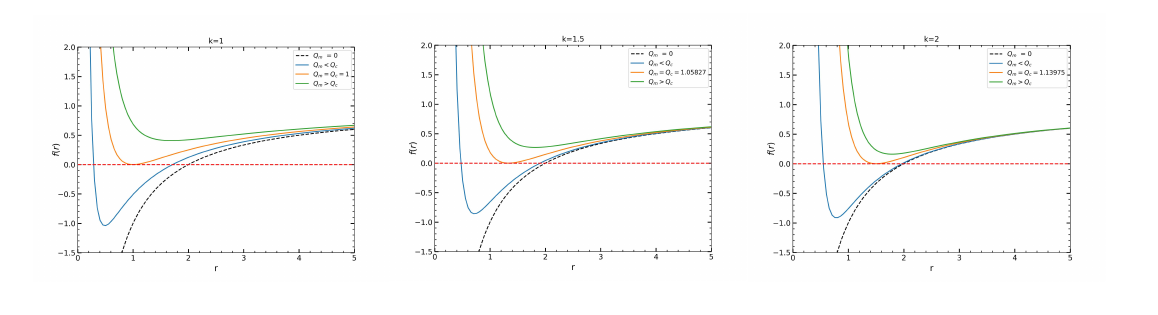}
\caption{
The existence of event horizons under different short hair BH backgrounds. From left to right are the short hair BH models with $k = 1$, $1.5$, and $2$ in turn.}
\label{a}
\end{figure*}

In order to explore the characteristics of particle orbital motion in the short hair BH spacetime, we will limit the range of short hair parameters to seek situations where there are no naked singularities. Therefore, the condition for an extremal BH is
\begin{equation}
\begin{cases}
g^{rr}=0 \\
\frac{dg^{rr}}{dr}|_{r = r_h}=0
\end{cases}.
\label{5}
\end{equation}
Bringing the short hair black hole (\ref{1}) into Eq. (\ref{5}) can be rewritten as
\begin{equation}
\begin{cases}
f\left(r\right)=0 \\
\left.\frac{df\left(r\right)}{dr}\right|_{r=r_h}=0
\end{cases}.
\label{6}
\end{equation}
Here, we stipulate that the hair parameter \(Q_m\) of the short hair BH reaches a critical value \(Q_c\), that is, \(Q_m = Q_c\). This condition corresponds to the situation of an extremal BH. Therefore, the value range of the hairy parameter is $0\leq Q_m\leq Q_c$.
As shown in Fig. \ref{a}, the change in the number of event horizons of the short hair BH when the hair parameter $Q_m$ takes different values. Obviously, when the hair parameter is $Q_c$, the short hair BH at this time is an extreme BH (the Cauchy horizon and the event horizon are degenerate). When the hair parameter $Q_m = 0$, the BH at this time degenerates into a Schwarzschild BH. The black dashed line in the figure represents the Schwarzschild BH case. It is easy to obtain from the figure that the existence of hair makes the event horizon radius smaller than the event horizon radius of the Schwarzschild BH. When the hair parameter takes a value of $Q_m < Q_c$, there are two horizons for the short hair BH at this time. When the hair parameter takes a value of $Q_m > Q_c$, there is no BH in the spacetime at this time, but a naked singularity appears. In this article, we focus on the case where the short hair BH always exists. Therefore, the value range of the hairy parameter is $0\leq Q_m\leq Q_c$.

For a compact object of stellar mass orbiting a supermassive compact object (here specifically referring to a BH), at this time, the binary system can be regarded as an extreme mass ratio system. The compact object of stellar mass can act as a test particle moving around the BH, and its motion follows the timelike geodesic equation. When the motion of the test particle is restricted to the equatorial plane ($\theta=\frac{\pi}{2}$), the Lagrangian can be written as
\begin{equation}
2\mathcal{L}=g_{\mu\nu}{\dot{x}}^\mu{\dot{x}}^\nu=-f\left(r\right){\dot{t}}^2+\frac{1}{f\left(r\right)}{\dot{x}}^2+r^2{\dot{\phi}}^2.
\label{7}
\end{equation}
Here, the overdot represents taking the derivative with respect to the affine parameter $\tau$, that is, ${\dot{x}}^\mu=\frac{dx^\mu}{d\tau}$. For a static spherically symmetric short hair BH, the corresponding expressions for the two conserved quantities, energy and angular momentum, can be written as
\begin{equation}
E=\frac{\partial\mathcal{L}}{\partial\dot{t}}=g_{tt}\dot{t}=f\left(r\right)\dot{t},
\label{8}
\end{equation}
and 
\begin{equation}
L=-\frac{\partial\mathcal{L}}{\partial\dot{\phi}}=-g_{\phi\phi}\dot{\phi}=-r^2\dot{\phi}.
\label{9}
\end{equation}

In general relativity, the four-velocity \({\dot{x}}^\mu\) of a particle moving along a timelike geodesic is a timelike unit vector. This means it satisfies the corresponding normalization condition, namely, that the four-velocity is normalized. The corresponding expression is
\begin{equation}
g_{\mu\nu}{\dot{x}}^\mu{\dot{x}}^\nu=-1.
\label{10}
\end{equation}
Bringing the metric (\ref{1}) and expressions (\ref{8}) and (\ref{9}) into Eq. (\ref{10}) and arranging yields
\begin{equation}
{\dot{r}}^2=E^2-V_{eff},
\label{11}
\end{equation}
where $V_{eff}$ represents the effective potential. The corresponding expression can be written as
\begin{equation}
V_{eff}=f\left(r\right)\left(1+\frac{L^2}{r^2}\right)=\left(1-\frac{2M}{r}+\frac{Q_m^{2k}}{r^{2k}}\right)\left(1+\frac{L^2}{r^2}\right).
\label{12}
\end{equation}
Obviously, it can be seen from the effective potential (\ref{12}) that when $r\rightarrow\infty$, at this time $\lim\limits_{r\rightarrow\infty}{V_{eff}} = 1$. In this case, if $E>1$, the particle can escape to infinity at this time ($\left.{\dot{r}}^2\right|_{r\rightarrow\infty}>0$), which is not what we are interested in. We are interested in the case where $E\leq1$ because in this case, due to the existence of the effective potential $V_{eff}$, the motion of the particle may be bound within a certain region. A detailed analysis will be carried out in the next subsection.

\section{\label{sec:level3}Bound Orbits of Test Particles in Short Hair Black Hole Spacetimes}
In this section, we will deeply explore two special types of orbits: the marginally bound orbit (MBO) and the innermost stable circular orbit (ISCO). The MBO corresponds to the critical bound state of particles in the gravitational field of a BH, with the maximum energy $E = 1$ at this time. The particles moving on this orbit are extremely unstable, and any tiny perturbation can cause the particle to deviate from the orbit and eventually fall into the BH or escape. The ISCO marks the innermost circular orbit where particles can stably exist. Any circular orbit inside the ISCO is unstable, and after being perturbed, the particle will inevitably spiral inward and finally fall into the BH.

\begin{figure*}[]
\includegraphics[width=1 \textwidth]{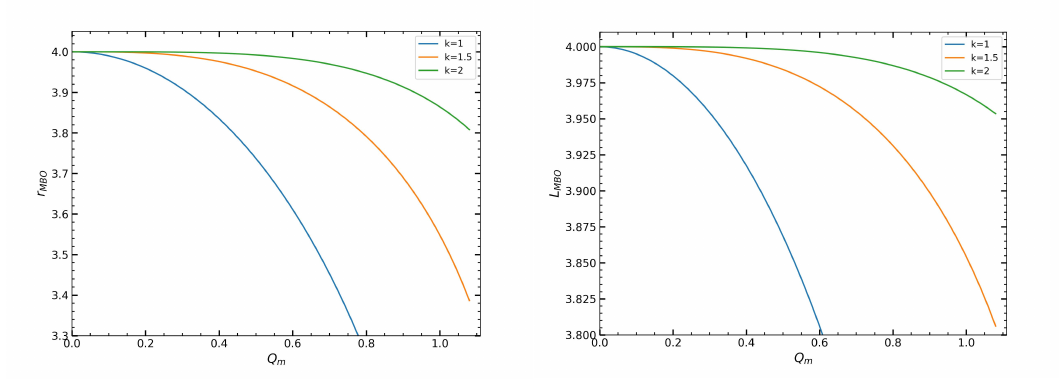}
\caption{
The variation of the radius $r_{MBO}$ and angular momentum $L_{MBO}$ of the MBO with the hairy parameter. The left figure shows the variation of the radius $r_{MBO}$ of the MBO with the hairy parameter. Curves of different colors represent different types of short hair BHs. From bottom to top, they represent $k = 1$, $1.5$, and $2$ respectively. The right figure represents the variation of the angular momentum $L_{MBO}$ of the MBO. Similarly, from bottom to top, they correspond to short hair BH types with $k = 1$, $0.5$, and $2$ respectively. Here $M = 1$.}
\label{b}
\end{figure*}

For the MBO, the conditions it satisfies are
\begin{equation}
V_{eff}=E^2=1,\quad \frac{dV_{eff}}{dr}=0.
\label{13}
\end{equation}
Obviously, it is very difficult to express the radius and angular momentum of the MBO in an analytically solved form. Therefore, it is very easy to obtain its corresponding changing trend by means of numerical solution here. As shown in Fig. \ref{b}, the radius $r_{MBO}$ and angular momentum $L_{MBO}$ of the MBO gradually decrease as the hairy parameter $Q_m$ increases. The smaller the value of parameter$k$, the more obvious the change. That is to say, the smaller the value of parameter $k$, the more significant the influence of the hairy parameter $Q_m$ on the MBO in this short hair BH. When the parameter $k$ takes a larger value, this trend will be suppressed. It is worth mentioning that when the parameter $k$ takes a value of 1, the short hair BH at this time degenerates into a Reissner-Nordstrom BH. Our numerical calculation results are consistent with the case of using neutral particles in the literature \cite{Misra:2010pu}. When the hairy parameter $Q_m$ disappears ($Q_m = 0$), the short hair BH at this time becomes a Schwarzschild BH. The radius $r_{MBO}=4M$ of the MBO is read from Fig. \ref{b}, which is consistent with the calculation result of the Schwarzschild BH. Our results show that the existence of the hairy parameter makes its orbital radius and angular momentum smaller than those of the Schwarzschild BH.

\begin{figure*}[]
\includegraphics[width=1 \textwidth]{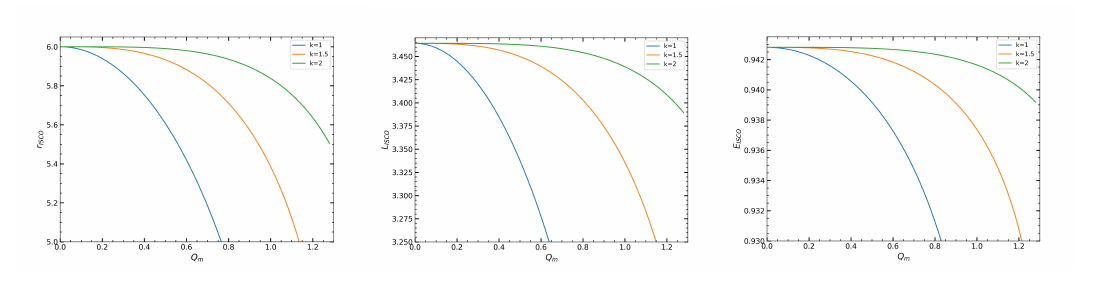}
\caption{
The variation of the ISCO radius $r_{ISCO}$, angular momentum $L_{ISCO}$, and energy $E_{ISCO}$ with the hairy parameter. Curves of different colors represent different types of short hair BHs. From bottom to top, they represent $k = 1$, $1.5$, and $2$ respectively.}
\label{c}
\end{figure*}

For the ISCO, the conditions it satisfies are
\begin{equation}
V_{eff}=E^2,\quad \frac{dV_{eff}}{dr}=0,\quad \frac{d^2V_{eff}}{dr^2}=0.
\label{14}
\end{equation}
Similarly, just like analyzing the MBO, the variation of the radius and angular momentum of the ISCO with the hairy parameter is obtained by means of numerical simulation. As shown in Fig. \ref{c}, the radius, angular momentum, and energy of the ISCO gradually decrease as the hairy parameter increases. When the value of parameter $k$ is small, the change is more drastic. When the parameter $k$ is large, it is suppressed. When the hairy parameter $Q_m$ disappears ($Q_m = 0$), the short hair BH degenerates into a Schwarzschild BH at this time. From Fig. \ref{c}, the radius $r_{ISCO}=6M$ of the ISCO can be read, which is in good agreement with the value of the Schwarzschild BH. Obviously, the existence of the hairy parameter makes the radius, angular momentum, and energy of the ISCO all lower than the corresponding values of the Schwarzschild BH. This means that particles will move along an orbit closer to the BH, thus providing us with an opportunity to more deeply detect and study the extreme gravitational field effects near BHs. In particular, the quantum effects caused by short hair parameters, because these quantum effects may carry information inside the event horizon.

\begin{figure}[]
\includegraphics[width=0.5 \textwidth]{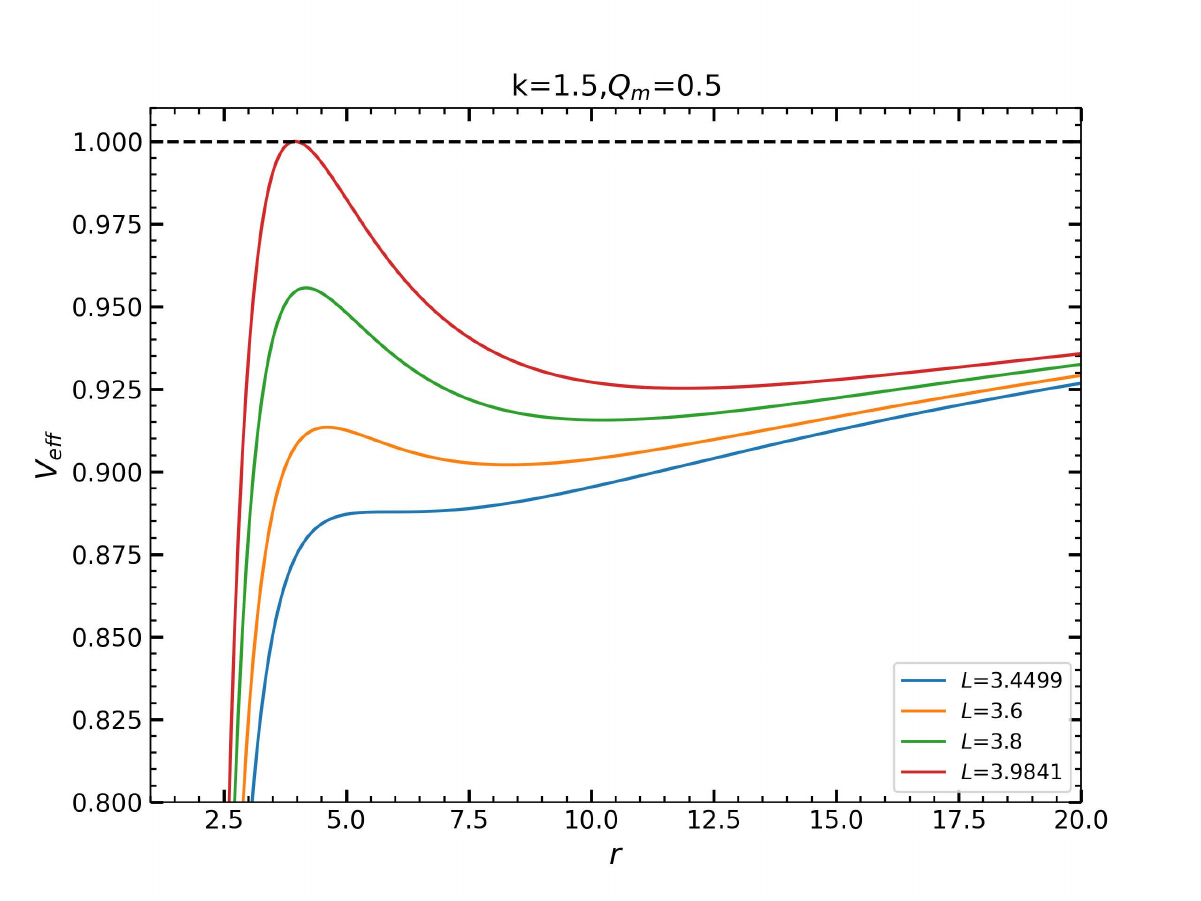}
\caption{
The variation of the effective potential with $r$ under different angular momenta. Here, the parameters $k = 1.5$ and the hairy parameter $Q_m = 0.5$ are fixed.}
\label{d}
\end{figure}

\begin{figure*}[]
\includegraphics[width=1 \textwidth]{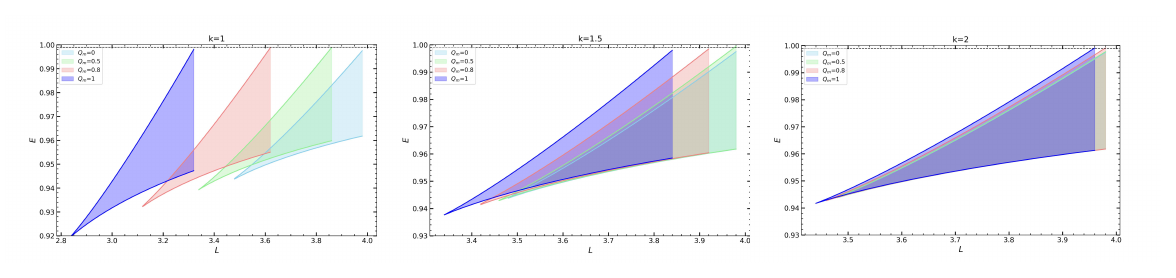}
\caption{
The $E-L$ regions allowed for particle orbits under different hairy parameters $Q_m$ and parameter $k$. Among them, from left to right, they correspond to parameter $k$ taking values of $1$, $1.5$, and $2$ respectively.}
\label{e}
\end{figure*}

When treating a compact object as a neutral test particle orbiting around a BH (In an EMR system.). If the particle's motion is located between the MBO and the ISCO, the bound state of the particle can be analyzed by means of the equation of motion of the test particle (see Eq. (\ref{11}) and the effective potential (\ref{12})).
As shown in Fig. \ref{d}, when the parameter $k$ is fixed at $1.5$ and the hairy parameter $Q_m = 0.5$, we plot the variation of the effective potential with the radius $r$. The red curve represents the situation of the MBO ($L_{MBO}=3.9841$), which has two extreme points. The blue curve represents the situation of the ISCO ($L_{ISCO}=3.4499$), which has only one extreme point. For the orbits between these two curves, the effective potential decreases as the angular momentum decreases, and the two extreme points keep approaching and will eventually converge at the ISCO.
In addition, for the bound orbit between the MBO and the ISCO, its energy will also be bound within the range of $E_{ISCO}<E<E_{MBO}$, and the size of the range can be determined by the angular momentum. As shown in Fig. \ref{e}, the $E-L$ space allowed for the particle's bound orbit under different parameters. Under the premise of maintaining generality, we have respectively plotted three short hair BH models ($k = 1, 1.5, 2$), and the influence of different hairy parameters $Q_m$ on the $E-L$ space of the particle's bound orbit.
Just as shown in Fig. \ref{e}, for the given parameter values, there is only one $E-L$ space corresponding to it. The smaller the parameter $k$, the more obvious the influence of the hairy parameter $Q_m$ on the $E-L$ space of its bound orbit. When the parameter $k$ takes a value of 1 and the hairy parameter $Q_m$ takes an extreme value, at this time, the influence of the hairy parameter on its energy-momentum space is the most obvious, and the deviation of its $E-L$ parameter space from the parameter space in the Schwarzschild BH case is more obvious (when the hairy parameter $Q_m = 0$, the short hair BH degenerates into a Schwarzschild BH. The space filled with light blue in Fig. \ref{e} represents the situation of the Schwarzschild BH). When the parameter $k$ takes a larger value, no matter how large the hairy parameter $Q_m$ is, the bound space that particles can satisfy will be degenerated to the case of Schwarzschild BH. This can be easily seen from Fig. \ref{e}. When the parameter $k$ takes a larger value, the $E-L$ spaces corresponding to different hairy parameters $Q_m$ will inevitably approach the situation of the Schwarzschild BH and finally degenerate together.

\begin{figure}[ht]
\includegraphics[width=0.5 \textwidth]{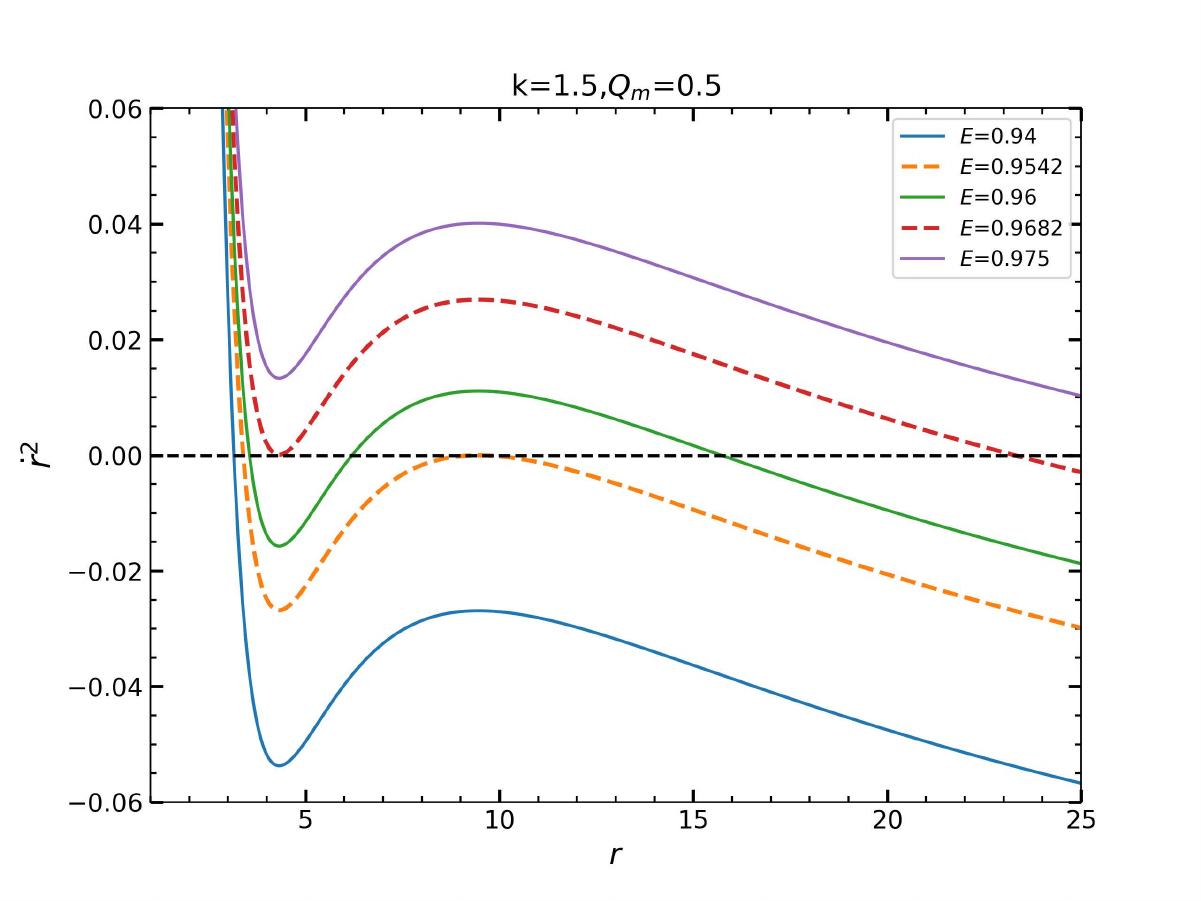}
\caption{
The function of $\dot{r}^2$, the equation of motion of particles around a short hair BH, with respect to the radial coordinate $r$. Here, with parameter $k$ fixed at $1.5$ and $Q_m = 0.5$, the existence of bound orbits under different energies is shown. Among them, $L=\frac{L_{ISCO}+L_{MBO}}{2}=3.717$.}
\label{f}
\end{figure}

In Fig. \ref{f}, we show the bound state of the test particle around the short hair BH model with fixed parameters $k = 1.5$ and hairy parameter $Q_m = 0.5$. Here, the angular momentum is fixed as $L=\frac{L_{ISCO}+L_{MBO}}{2}=3.717$.
Obviously, in the process of the energy $E$ gradually increasing, the change of the root of the equation of motion $\dot{r}^2 = 0$ is as follows: there is one root, two roots, three roots, two roots, and finally one root. This can be physically described as whether the particle motion is bound.
When $E<0.9542$, there is only one root at this time. In this case, the particle can only spiral into the BH at a specific position. When $E = 0.9542$ (orange dotted line in the figure), this is a critical value and there are two roots. But it is easy to see from the figure that there is $\dot{r}^2<0$ between the two roots, which also means that there is no bound orbit in this case. When the energy range is within $0.9542<E<0.9682$, there are three roots at this time. It is easy to see from the figure that there is $\dot{r}^2>0$ between the latter two roots, which indicates that there are bound orbits for the particle motion around the short hair BH within this range. This is in agreement with the numerical value of the $E-L$ space given in Fig. \ref{e}. When $E = 0.9682$ (red dotted line in the figure), there are two roots. There is $\dot{r}^2>0$ between these two roots, which indicates that there are bound orbits in this case. However, when the energy $E>0.9682$, there is only one root at this time, and it can be seen from the figure that $\dot{r}^2>0$ within a limited range, which indicates that the particle can spiral and fall into the BH from a finite distance.

\section{\label{sec:level4}Periodic Orbit Dynamics in an  Extreme Mass Ratio System under the Background of a Short Hair Black Hole}
\begin{figure*}[]
\includegraphics[width=1 \textwidth]{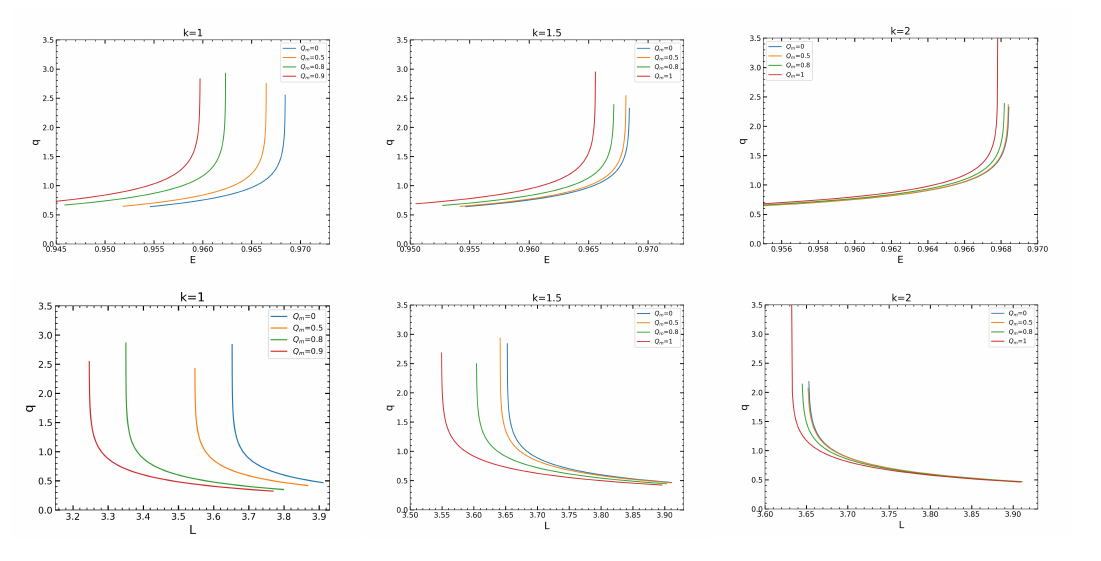}
\caption{
The variation of rational number $q$ with $E$ or $L$. In the first row, with $L=\frac{L_{ISCO}+L_{MBO}}{2}$, the graph of $q$ versus $E$ is plotted. In the second row, with $E = 0.96$, the graph of $q$ vs $L$ is plotted. From left to right are the short hair BH models representing $k = 1$, $1.5$, and $2$ respectively.}
\label{g}
\end{figure*}

In the context of a short hair BH spacetime, periodic orbits in an EMR system exhibit a range of unique physical characteristics. The orbital structure and evolutionary behavior of stellar-mass compact objects, influenced by the strong gravitational field of the primary BH, differ significantly from those in conventional mass ratio systems. Notably, in the modified spacetime background of a short hair BH, these orbits may display periodic behaviors and orbital evolutions that are markedly different from those predicted by classical BH models, thereby having significant implications for the overall dynamics of the system and the characteristics of gravitational wave radiation. In this section, we will explore the properties of these periodic orbits. We adopt the classification scheme from the literature \cite{Levin:2008mq} to describe these orbits $(z, w, v)$ using a unique irrational or rational number \( q \). The expression is given by
\begin{equation}
q = \frac{\Delta \phi}{2\pi} - 1 = w + \frac{v}{z}.
\label{15}
\end{equation}
In expression (\ref{15}), \( w \) represents the whirl number of the orbit, \( v \) denotes the vertex number of the orbit, and \( z \) signifies the zoom number (or blade number). It is worth noting that these three elements are all integers.
When $q$ is an irrational number, the orbit of a neutral particle at this time exhibits precession phenomena, similar to the perihelion precession of Mercury. When $q$ is a rational number, the motion of the particle at this time shows periodic characteristics, namely a periodic orbit.
As pointed out in reference \cite{Levin:2008mq}, any general orbit can be approximated by a periodic orbit. Therefore, the study of periodic orbits is helpful for understanding the properties of the BH background and has potential application value in understanding the gravitational radiation mechanism around BHs.
$\Delta\phi$ represents the angular change within each period. Its expression is
\begin{equation}
\Delta\phi=\oint d\phi = 2\int_{r_1}^{r_2}\frac{\mathrm{d}\phi}{\mathrm{d}r}\mathrm{d}r,
\label{16}
\end{equation}
where $r_1$ and $r_2$ are two turning points of the particle motion and are determined by the equation of motion (\ref{11}), that is, the latter two roots of ${\dot{r}}^2 = 0$. Combining Eqs. (\ref{9}), (\ref{11}) and (\ref{16}), Eq. (\ref{15}) can be written as
\begin{align}
q&=\frac{\Delta \phi}{2\pi}-1=\frac{1}{\pi}\int_{r_1}^{r_2}\frac{d \phi}{d r} dr-1 \nonumber\\
&=\frac{1}{\pi}\int_{r_1}^{r_2}\frac{L}{r^2\sqrt{E^2-\left(1-\frac{2M}{r}+\frac{Q_m^{2k}}{r^{2k}}\right)\left(1+\frac{L^2}{r^2}\right)}} dr-1.
\label{17}
\end{align}
Here we use numerical methods to plot the variation of the rational number \(q\) with respect to \(E\) or \(L\). As shown in Fig. \ref{g}. In the first row of figures, since the orbital angular momentum range is $L_{ISCO}<L<L_{MBO}$, the angular momentum $L$ can be expressed as $L = L_{ISCO}+\varepsilon(L_{MBO}-L_{ISCO})$. Without loss of generality, we fix $\varepsilon=\frac{1}{2}$ to plot the curve of $q$ varying with E (see Fig. \ref{g}).
Apparently, within the energy range of $E_{ISCO}<E<E_{MBO}$, the rational or irrational number $q$ under the background of three types of short hair BHs starts to increase slowly with the increase of energy and rises sharply to divergence near the maximum energy value. The existence of hair causes the curve to shift to the left and diverge at the maximum value of its corresponding energy, and the greater the hair strength, the faster it will reach the divergence point. This is consistent with the transformation situation of the $E - L$ space shown in Fig. \ref{e}. That is to say, the existence of hair makes it diverge earlier compared with the Schwarzschild BH case, and the greater the hair strength, the faster the divergence. The blue curve in the figure represents the situation of the Schwarzschild BH.
It is worth mentioning that we find that when the parameter $k$ increases, the behavior of the hair parameter on $q$ will approach the situation of the Schwarzschild BH and eventually degenerate when $k$ takes a larger value. In the second row of figures, we fix the energy $E = 0.96$ to plot the variation of $q$ with the angular momentum $L$. Apparently, as the angular momentum decreases, $q$ first increases slowly. When the angular momentum decreases to near its minimum value, $q$ increases sharply and eventually diverges. The existence of hair parameters makes it diverge earlier, and the stronger the intensity, the more obvious it is. As the value of $k$ increases, these trends will approach the situation of the Schwarzschild BH and eventually degenerate (the blue curve in the figure represents the situation of the Schwarzschild BH).

For the trajectories of the periodic orbits $(z, w, v)$ we are concerned about, the energies and angular momenta corresponding to different orbits are given in Tables \ref{table1} and \ref{table2}. Obviously, for a fixed angular momentum $L=\frac{L_{ISCO}+L_{MBO}}{2}$, the energy $E$ per period decreases as the hair parameter $Q_m$ increases and increases as the parameter $k$ increases. In addition, for orbits with the same period, compared with classical BHs, the periodic orbit energy $E$ of short hair BHs is more significantly dependent on hair parameters. That is, the larger the hair parameter, the more obvious the difference in energy. And when the parameter $k$ is smaller, this difference is also more obvious (see Table \ref{table1}). For a fixed energy $E = 0.96$, the angular momentum $L$ of the same periodic orbit decreases as the hair parameter increases and increases as the parameter $k$ increases (see Table \ref{table2}). In general, compared with the classical Schwarzschild BH case, the periodic orbits of short hair BHs have smaller energy and angular momentum.

\begin{table*}[]
\centering
\begin{tabular}{p{1.5cm}p{1.5cm}p{2cm}p{2cm}p{2cm}p{2cm}p{2cm}p{2cm}}
\hline\hline
\rule{0pt}{12pt}
$k$& $Q_m$& $E_{(1,1,0)}$& $E_{(1,2,0)}$ & $E_{(2,1,1)}$ & $E_{(2,2,1)}$ & $E_{(3,1,2)}$ & $E_{(3,2,2)}$ \\
\hline
\rule{0pt}{11pt} 
1&	0&	0.96542534& 0.96838276& 0.96802649& 0.96843430& 0.96822485 &	0.96843848 \\
\rule{0pt}{11pt}
&	0.5&	0.96313564& 0.96643290&	0.96602526&	0.96649366&	0.96625089&	0.96649857  \\
\rule{0pt}{11pt}
&	0.8&0.95793699&	0.96222322&	0.96165488& 0.96231514&	0.96196436&	0.96232330\\ 
\rule{0pt}{11pt}
&	0.9& 0.95443178 &0.95958114 &0.95885940 &0.95970575&	0.95924720&	0.95971742 \\ 
\hline
\rule{0pt}{11pt}
1.5&0&0.96542534 &0.96838276&	0.96802649&	0.96843430&	0.96822485&	0.96843848\\
\rule{0pt}{11pt}
&0.5&0.96501094&	0.96808418&0.96770971&	0.96813908&0.96791765 &0.96814358 \\
\rule{0pt}{11pt}
&0.8&0.96353867&	0.96705669&	0.96660965&	0.96712555&	0.96685548&	0.96713143\\
\rule{0pt}{11pt}
&1&0.96106760&	0.96545151&	0.96485064&	0.96555276&	0.96517515&	0.96556207 \\
\hline
\rule{0pt}{11pt}
2&	0&0.96542534 &0.96838276&	0.96802649& 	0.96843430& 0.96822485&	0.96843848\\ 
\rule{0pt}{11pt}
&0.5&0.96536321 &0.96834518&	0.96798484 &0.96839749&0.96818532&	0.96840175 \\
\rule{0pt}{11pt}
&0.8&0.96500258 &0.96812981& 	0.96774510 &0.96818685 &0.96795826&0.96819157 \\
\rule{0pt}{11pt}
&1&0.96431862 &0.96773424 &	0.96729929&	0.96780141&	0.96753836 &0.96780716 \\
\hline\hline
\end{tabular}
\caption{The energy corresponding to the periodic orbit $(z, w, v)$ under different BH models and different hair intensities. Here, the angular momentum is fixed as $L=\frac{L_{ISCO}+L_{MBO}}{2}$.}
\label{table1}
\end{table*}

\begin{table*}[]
\centering
\begin{tabular}{p{1.5cm}p{1.5cm}p{2cm}p{2cm}p{2cm}p{2cm}p{2cm}p{2cm}}
\hline\hline
\rule{0pt}{12pt}
$k$& $Q_m$& $L_{(1,1,0)}$& $L_{(1,2,0)}$ & $L_{(2,1,1)}$ & $L_{(2,2,1)}$ & $L_{(3,1,2)}$ & $L_{(3,2,2)}$ \\
\hline
\rule{0pt}{11pt} 
1&0&3.68358773 &	3.65340563 &3.65759566&	3.65270065&3.65533454&3.65263628 \\
\rule{0pt}{11pt}
&0.5&3.57720722 &3.54743482 &3.55152187 &	3.54675867 &3.54930910&	3.54669780 \\
\rule{0pt}{11pt}
&0.8&3.38254497 &3.35143400 &3.35576865 &3.35070702&3.35342892 &3.35064092 \\
\rule{0pt}{11pt}
&0.9&3.28087279&3.24746356&3.25227471&	3.24662491&	3.24969858&3.24654628 \\
\hline
\rule{0pt}{11pt}
1.5&0&3.68358773&3.65340563&3.65759566&	3.65270065&	3.65533454&	3.65263628 \\
\rule{0pt}{11pt}
&0.5&3.67287100&3.64223818&	3.64650849&	3.64151645&	3.64420624&3.64145032\\
\rule{0pt}{11pt}
&0.8&3.63744919&3.60499338&	3.60960574&3.60419669&	3.60713041&	3.60412243 \\
\rule{0pt}{11pt}
&1&3.58671507&3.55048399&3.55587308&3.54950125&	3.55301361&3.54940561 \\
\hline
\rule{0pt}{11pt}
2&0	&3.68358773&3.65340563&3.65759566&3.65270065&3.65533454 &3.65263628 \\
\rule{0pt}{11pt}
&0.5&3.68251759&3.65220156&3.65641802 &3.65149058 &3.65414366 &3.65142554 \\
\rule{0pt}{11pt}
&0.8&	3.67645938&3.64535200&3.64972680 &3.64460460&3.64737342&	3.64453551\\
 \rule{0pt}{11pt}
&1&3.66567156&3.63300137&3.63769976&3.63217695 &	3.63518629&3.63209910 \\
\hline\hline
\end{tabular}
\caption{The angular momentum corresponding to the periodic orbit $(z, w, v)$ under different BH models and different hair intensities. Here, the energy is fixed as $E = 0.96$.}
\label{table2}
\end{table*}

\begin{figure*}[ht]
\includegraphics[width=1 \textwidth]{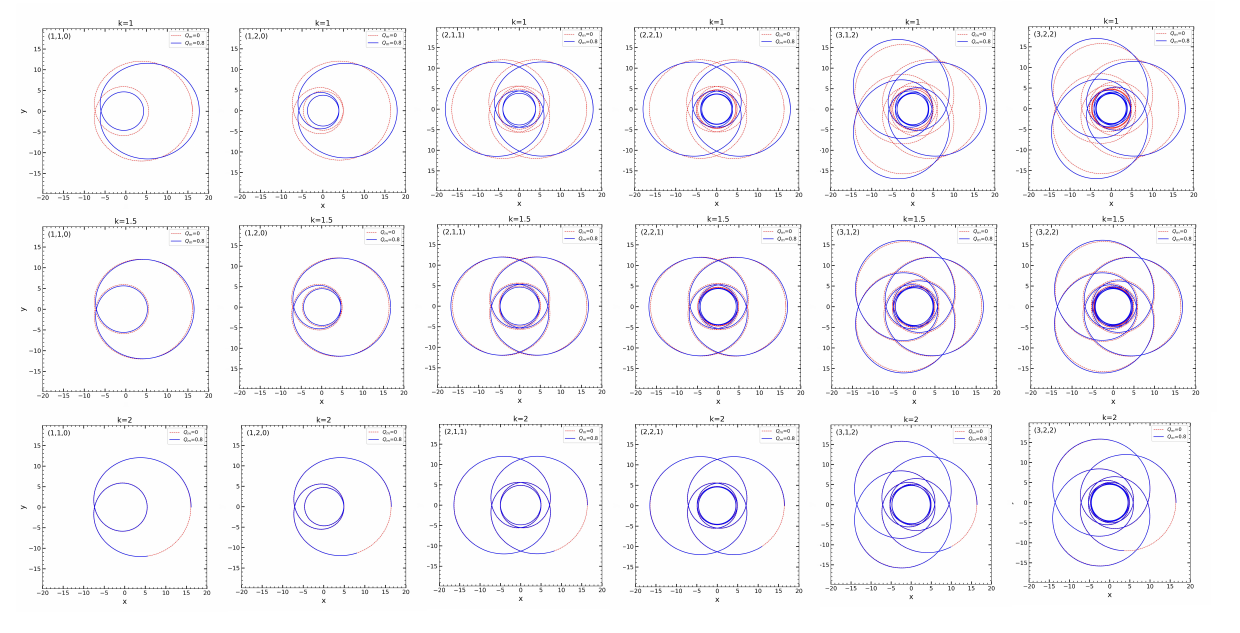}
\caption{
The trajectory of the periodic orbit $(z, w, v)$ of a neutral test particle in the spacetime background of the short hair BH. The figures from top to bottom represent the short hair BH models with parameter $k$ taking values of $1$, $1.5$, and $2$ in turn. Here, we have fixed the hair parameter $Q_m = 0.8$, energy $E = 0.96$, and the angular momentum can be obtained from Table \ref{table2}. Among them, the red curve represents the periodic orbit of the Schwarzschild BH, and the blue curve represents the periodic orbit of a short hair BH. Here, the coordinate transformation is $(x,y)=(r\cos\phi,r\sin\phi)$.}
\label{h}
\end{figure*}

\begin{figure*}[ht]
\includegraphics[width=1 \textwidth]{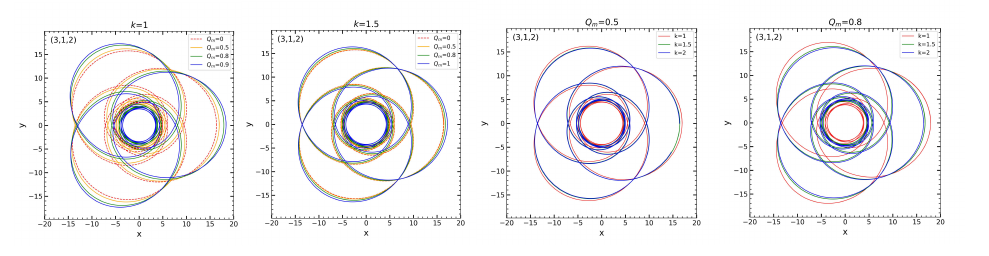}
\caption{
The trajectories of the periodic orbit $(3, 1, 2)$ under different hair parameters $Q_m$ or parameter $k$. The energy is fixed at $0.96$, and the angular momentum can be obtained from Table \ref{table2}. Here, the coordinate transformation is $(x,y)=(r\cos\phi,r\sin\phi)$.}
\label{i}
\end{figure*}

In Figs. \ref{h} and \ref{i}, we plot the trajectory diagrams of the periodic orbits $(z, w, v)$ in the spacetime background of short hair BHs according to the data in Table \ref{table2}. Here, $E = 0.96$ and $Q_m = 0.8$ are fixed, and the angular momentum is read from Table \ref{table2}.
Obviously, as pointed out in reference \cite{Levin:2008mq}, the parameter $z$ describes the zoom number of the periodic orbit. $w$ represents the whirl number generated when a particle approaches a short hair BH. $v$ represents the vertex number of the running sequence in the particle trajectory. These parameters jointly characterize the motion characteristics of particles in a strong gravitational field. As the zoom number $z$ increases, the complexity of the particle's trajectory also increases. These three parameters reflect the richer dynamical behavior of EMR systems within the framework of general relativity.
In Fig. \ref{h}, we plot the periodic orbits under the Schwarzschild BH and the short hair BH with the same trajectory $(z, w, v)$. Obviously, due to the existence of the hair parameter $Q_m$, the outermost trajectory of the periodic orbit becomes larger, while the inner trajectory is closer to the BH. When the parameter $k$ takes a relatively large value, the trajectory of the short hair BH will gradually approach that of the Schwarzschild BH and eventually degenerate (see Fig. \ref{h}). The red dashed line in the figure represents the case where the short hair BH degenerates into the Schwarzschild BH ($Q_m = 0$).
In Fig. \ref{i}, we show the influence of hair parameters or parameter $k$ on the periodic orbit under the same trajectory $(z, w, v)$. It can be easily seen from the figure that the larger the hair parameter $Q_m$ is, the more obvious the difference between the particle trajectory in the spacetime background of the short hair BH and that in the case of the Schwarzschild BH. And this difference makes the outermost orbit extend outward. The larger the parameter $k$ is, the orbit will eventually degenerate (see Fig. \ref{i}).

In general, although the innermost orbits of these periodic orbits are not located near the event horizon radius of the short hair BH and do not even break through the boundary of the photon ring radius, the changes in the spacetime structure near the event horizon caused by the hair parameters of the short hair BH still have an indirect and significant impact on these periodic orbits (which can be clearly seen from the periodic orbits). Compared with the Schwarzschild BH, the distortion of the spacetime geometry near the event horizon of the short hair BH accumulates over a greater distance, resulting in the periodic orbits exhibiting an outward expansion characteristic. This phenomenon reveals the long-range effect of the short hair parameters on the external spacetime of the BH. This not only helps to distinguish between the short hair BH and the Schwarzschild BH but also provides a new observational window for testing the no-hair theorem and even the "no-short-hair" theorem. In addition, this long-range effect may be related to quantum effects, further enriching the complexity of orbital dynamics. 
Therefore, through precise measurements of these periodic orbits, it is possible to capture the unique information of the short hair BH both experimentally and observationally, thus providing new evidence for the research of general relativity and quantum gravity.

\section{\label{sec:5}Gravitational Wave Radiation of Periodic Orbits in an Extreme Mass Ratio System under Short Hair Black Hole Background}
In an EMR system, a short hair BH plays the role of a massive BH. A small-mass celestial body undergoes periodic motion in its strong gravitational field and may radiate gravitational waves. Since the mass of the small-mass celestial body is much smaller than that of the short hair BH, its perturbation to the background spacetime can be regarded as negligible. In this case, it is reasonable to adopt the adiabatic approximation, that is, within one or several orbital periods, the energy and angular momentum losses of the system can be ignored.

Under the adiabatic approximation, we assume that the energy and angular momentum of the system remain unchanged within a single period. Thus, these conserved quantities (energy and angular momentum) can be regarded as fixed values to calculate the gravitational wave radiation within one period. This treatment method is effective in EMR systems. It can accurately and quickly describe the orbital evolution of a small celestial body in the background of a short hair BH and its gravitational wave radiation characteristics. This method has been well applied in references \cite{Tu:2023xab,Li:2024tld,Zi:2024dpi,Zi:2023qfk,Glampedakis:2002ya}.

To obtain the expression form of the gravitational wave signal, we adopt the kludge waveform developed in reference \cite{Babak:2006uv}. This method first determines the periodic orbital behavior of the test particle through the geodesic equation (the detailed discussion has been introduced in the previous section), and then uses the quadrupole formula to obtain the gravitational wave waveform. For the gravitational wave waveform radiated by a periodic orbital motion, if the second-order term is calculated, the quadrupole formula can be written as \cite{Maselli:2021men,Liang:2022gdk}
\begin{equation}
h_{ij}=\frac{4\mu M}{D_L}\left(v_iv_j-\frac{m}{r}n_in_j\right).
\label{18}
\end{equation}
In Eq. (\ref{18}), \(M\) represents the mass of the supermassive BH (here, the mass of the short hair BH), and \(m\) is the mass of the small celestial body. Considering that we are studying an EMR system, thus \(m\ll M\). \(D_L\) is the luminosity distance of the EMR system. \(\mu=\frac{Mm}{(M + m)^2}\) is the symmetric mass ratio of the system. \(v_i\) represents the component of the velocity vector of the small celestial body and describes the motion state of the small celestial body on its orbit; \(n_i\) is the unit vector pointing in the radial direction, that is, the direction vector pointing from the BH to the small celestial body, which is obtained after normalization processing. 

The quadrupole moment formula (\ref{18}) describes the spacetime perturbation of the gravitational wave generated by the motion of a small celestial body in the gravitational field of a supermassive BH. This waveform provides the basic distribution characteristics of the gravitational wave in space. However, in order to match it with the actual detector, we need to match the calculated gravitational wave signal with the actual detector, which requires projecting the gravitational wave onto the coordinate system suitable for the detector. As shown in the literature \cite{Poisson_Will_2014}, its coordinate basis can be written as
\begin{align}
e_X&=\left[\cos{\zeta},-\sin{\zeta},0\right],\nonumber\\
e_Y&=\left[\cos{\iota}\sin{\zeta},\cos{\iota}\cos{\zeta},-\sin{\iota}\right],\nonumber\\
e_Z&=\left[\sin{\iota}\sin{\zeta},\sin{\iota}\cos{\zeta},\cos{\iota}\right].
\label{19}
\end{align}
In the above coordinate basis, \(\zeta\) is the longitude of periastron, \(\iota\) is the inclination angle of the test particle orbit relative to the observation direction, and \(e_X\), \(e_Y\), \(e_Z\) define the orthogonal coordinate basis suitable for the detector and are used to decompose the gravitational wave signal into different polarization components. Under such a coordinate basis, the polarization components \(h_+\) and \(h_\times\) can be expressed as \cite{Poisson_Will_2014}
\begin{equation}
h_+=-\frac{2\mu M^2}{D_Lr}\left(1+\cos^2{\iota}\right)\cos{\left(2\phi+2\zeta\right)},
\label{20}
\end{equation}
and
\begin{equation}
h_\times=-\frac{4\mu M^2}{D_Lr}\cos{\iota}\sin{\left(2\phi+2\zeta\right)},
\label{21}
\end{equation}
here, $\phi$ represents the phase angle.

\begin{figure*}[ht]
\includegraphics[width=1 \textwidth]{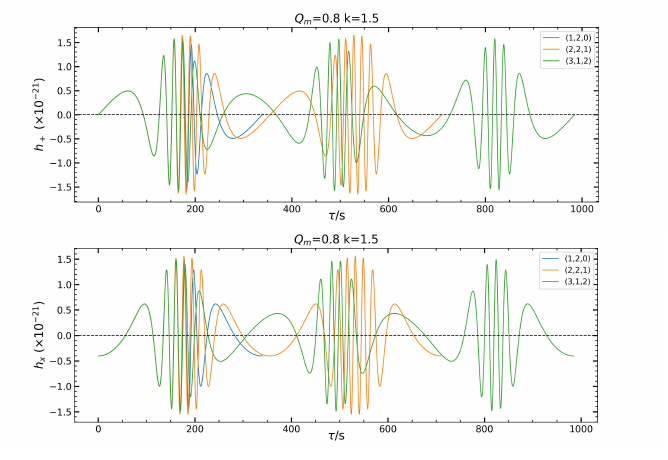}
\caption{
The gravitational wave waveforms produced by different periodic orbits. The upper and lower figures correspond to gravitational waves with different polarization states respectively. Among them, the upper figure is for \(h_+\) and the lower figure is for \(h_{\times}\). The parameter values of the short hair BH model are taken as \(k = 1.5\) and \(Q_m = 0.8\).}
\label{j}
\end{figure*}

\begin{figure*}[ht]
\includegraphics[width=1 \textwidth]{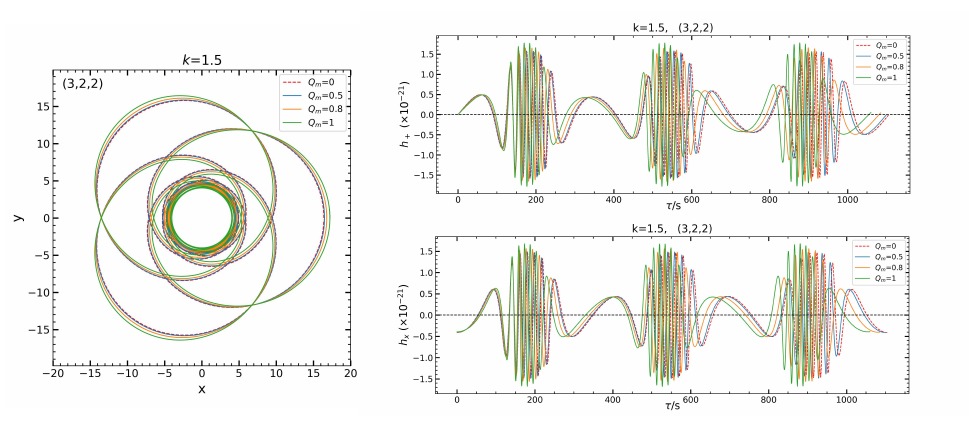}
\caption{
The gravitational wave waveforms corresponding to different hair parameters in the periodic orbit $(3, 2, 2)$. The left image represents the periodic orbits under different hair parameters, and the right image shows the gravitational wave waveforms radiated by the periodic orbits. The upper and lower figures correspond to different polarization directions respectively. At this time, the parameter value of the short hair BH model is \(k = 1.5\). When the hair parameter \(Q_m = 0\), the short hair BH degenerates into a Schwarzschild BH.}
\label{k}
\end{figure*}

\begin{figure*}[ht]
\includegraphics[width=1 \textwidth]{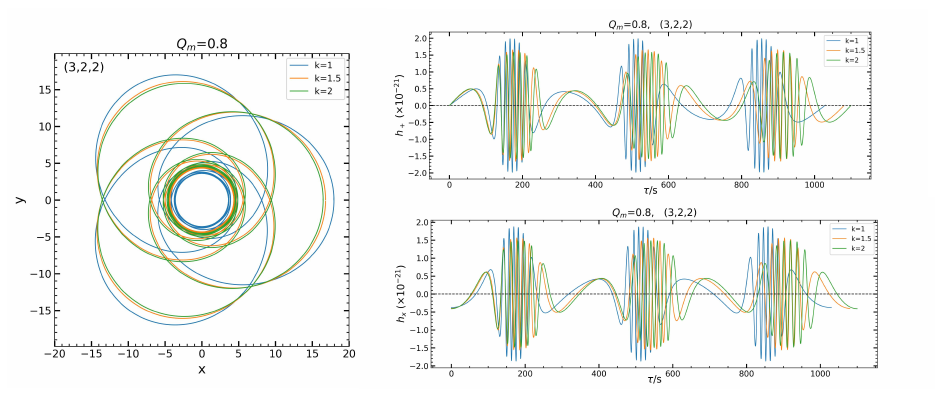}
\caption{
The gravitational wave waveforms radiated by different \(k\) values in the periodic orbit $(3, 2, 2)$. The left image represents the periodic orbits under different \(k\) parameters, and the right image shows the gravitational wave waveforms radiated by the periodic orbits. The upper and lower figures correspond to different polarization directions respectively. Among them, the parameter value of the short hair BH model is \(Q_m = 0.8\).}
\label{l}
\end{figure*}

To explore the influence of different parameter conditions (hair parameter \(Q_m\) or parameter \(k\)) on the gravitational waves radiated by small celestial bodies in periodic orbital motions. We assume the parameters of the EMR system are as follows: the mass of the short hair BH playing the role of the supermassive BH in the extreme mass ratio system is \(M = 10^7M_{\odot}\), and the mass of the small celestial body playing the role of the test particle is \(m = 10M_{\odot}\). As for the inclination angle \(\iota\) and the latitude angle \(\zeta\), they are set as \(\iota=\frac{\pi}{4}\) and \(\zeta=\frac{\pi}{4}\) respectively, and the luminosity distance \(D_L\) is set as \(D_L = 200\) Mpc.
Figs. \ref{j}, \ref{k}, and \ref{l} display the gravitational wave waveforms radiated by typical periodic orbits under different parameter settings. In Fig. \ref{j}, it can be readily observed that within a complete period, the gravitational wave waveforms (\(h_+\) and \(h_-\)) have corresponding zoom and whirl stages. The calm part of the waveform corresponds to the zoom stage, that is, when the small celestial body moves to the highly elliptical stage far from the BH. The part of the waveform with sharp oscillations corresponds to the whirl stage, that is, when the small celestial body approaches the short hair BH and exhibits circular whirl behavior. At this time, the gravitational wave frequency is relatively high and the oscillations are intense.
From the waveform of one period, the zoom-whirl behavior clearly shows the motion characteristics of the small celestial body within one period (the number of its calm phases corresponds to the number of blades of the small celestial body's periodic orbit, and the number of phases with sharp oscillations corresponds to the number of whirls of the periodic orbit).
To explore the influence of the hair parameter \(Q_m\) and parameter \(k\) on its gravitational waves, the waveforms of gravitational waves under different parameters are plotted in Figs. \ref{k} and \ref{l}. In Fig. \ref{k}, with the parameter \(k = 1.5\) and the orbital mode $(3, 2, 2)$ fixed, the influence of the hair parameter on its gravitational waves is plotted. Obviously, compared with the Schwarzschild BH case, a larger hair parameter changes the phase of the gravitational wave signal, reduces its period, and increases its amplitude (the red dashed line in the figure represents the Schwarzschild BH case). In Fig. \ref{l}, with the hair parameter \(Q_m = 0.8\) and the orbital mode $(3, 2, 2)$ fixed, the differences in the gravitational wave waveforms under three short hair BH models are plotted. The results show that a smaller value of \(k\) has a similar effect on the gravitational wave signal as a larger hair parameter.

In general, the hair parameter \(Q_m\) and parameter \(k\) exert a remarkable influence on the gravitational wave signals radiated by the periodic orbits. Specifically, a larger value of the hair parameter \(Q_m\) and a smaller value of parameter \(k\) will suppress the period of the gravitational wave signal, simultaneously increase the amplitude, and alter its phase. This implies that by identifying these differences, future gravitational wave detections can disclose the features of the zoom and whirl behaviors of the orbit. These features not only offer crucial evidence for identifying the properties of the short hair BH, distinguishing it from the Schwarzschild BH, but also open a potential observational window for verifying the no-hair theorem or the "no-short-hair" theorem.

It is worth noting that this paper employs the adiabatic approximation method to calculate the gravitational wave waveform. Although this method is unable to fully capture all the details of the gravitational wave, especially in terms of the contributions of high-order effects and non-adiabatic effects. However, it can effectively reveal the main physical characteristics of the gravitational wave signal within a single period. Especially, it is prominently manifested in describing the zoom and whirl behaviors of the periodic orbit, the influence of the hair parameter on the gravitational wave signal, and the differences from the waveform of the Schwarzschild BH.
Therefore, within the scope of this study, the adiabatic approximation is reasonable and sufficient (because this paper mainly explores whether the gravitational wave signal in a single period can exhibit basic orbital characteristics such as zoom and whirl behaviors). However, the adiabatic approximation ignores the contributions of non-adiabatic effects and high-order multipole moments in the long-term orbital evolution. Specifically, non-adiabatic effects may significantly affect the amplitude and phase of the waveform on a longer time scale, especially during the long-term orbital evolution. In addition, high-order effects may also change the details of the waveform, and these effects have not been fully considered in the current adiabatic approximation.

Therefore, to depict the evolution of the gravitational wave waveform more accurately, future research will adopt more refined waveform models, especially those that can simultaneously consider high-order effects and the contributions of non-adiabatic radiation. These improvements will not only deepen our understanding of the radiation from periodic orbits but also contribute to more accurately distinguishing the properties of the short hair BH from those of the classical Schwarzschild BH. Future space-based gravitational wave detectors (such as LISA, Tianqin, Taiji, etc.) will provide crucial observational data for verifying these theoretical models. For example, by studying the gravitational wave waveform in the last year of orbital inspiral through the LISA detector and combining it with the power spectral density($S_n$) of the detector, it can be used to explore the influence of the hair parameter $Q_m$ on the long-term evolution of periodic orbits or to place constraints on it, thereby enhancing our understanding of the characteristics of the short hair BH. These issues will be further explored in future research.

\section{\label{sec:6}Discussion and conclusions}
This paper discusses a short hair BH that can bypass the "no-short-hair" theorem \cite{Nunez:1996xv,Acharya:2024kvv}. Its remarkable feature is that it clearly exhibits the behavior of the hair structure near the event horizon. Under normal circumstances, this region will show strong quantum effects, and these effects may carry important information about the internal structure and evolutionary history of the black hole. For this reason, in an extreme mass ratio system, we regard the short hair BH as a supermassive BH and the small celestial body as a test particle, and study the characteristics of gravitational waves radiated by timelike geodesics and periodic orbital motions. The significant influence of hair parameters on these characteristics is analyzed emphatically. The main findings are as follows:

We have derived the general characteristics of the bound orbits of test particles in the background of a short hair BH. The results show that the hair parameter \(Q_m\) and parameter \(k\) have a significant influence on the radii and angular momenta of the MBO and the ISCO.
Specifically, as the hair parameter \(Q_m\) increases, the radii and angular momenta of these orbits gradually decrease. While an increase in parameter \(k\) will weaken the influence of \(Q_m\) on orbital changes. In other words, when parameter \(k\) takes a relatively large value, the short hair BH will tend to degenerate into a Schwarzschild BH (see Figs. \ref{b} and \ref{c}).
In addition, the hair parameter \(Q_m\) and parameter \(k\) also have an impact on the $E-L$ space allowed by the bound orbit. In particular, when \(k\) is small or the hair parameter \(Q_m\) takes extreme values, the $E-L$ space will shift significantly to the left. That is, the existence of hair parameters makes this space gradually deviate from the case of the Schwarzschild BH. As the parameter \(k\) increases, the $E-L$ space of the bound orbit of test particles in the background of the short hair BH will gradually approach the case of the Schwarzschild BH and eventually tend to degenerate (see Fig. \ref{e}).

Based on the $E-L$ space allowed by the bound orbit of test particles, we use the orbital classification method to encode periodic orbits \cite{Levin:2008mq}. We find that as the energy increases, the rational number \(q\) of the orbit slowly increases until a sharp increase occurs. Similarly, as the angular momentum decreases, \(q\) will also show a similar growth trend and eventually increase sharply. Compared with the classical Schwarzschild BH, both the hair parameter \(Q_m\) and parameter \(k\) have an impact on the orbit, resulting in the test particle around the short hair BH having lower energy and angular momentum (see Fig. \ref{g} and Tables \ref{table1} and \ref{table2}).
To visually demonstrate these differences, we have plotted several typical periodic orbits in Figs. \ref{h} and \ref{i}. Obviously, the existence of the hair parameter \(Q_m\) and the change of parameter \(k\) will significantly affect the periodic orbits. Especially in the case of larger \(Q_m\) and smaller \(k\), this influence is more obvious. This provides the possibility to distinguish the short hair BH from Schwarzschild BH. In addition, since the hair parameter will affect the spacetime structure near the event horizon and is related to quantum effects, these effects will accumulate on the periodic orbits and cause significant changes in their characteristics. Therefore, observing these orbital characteristics can not only reveal the internal structure and evolution of short hair BH but also provide a key basis for testing the no-hair theorem and its extensions (such as the "no-short-hair" theorem). This research may provide a new perspective and support for exploring general relativity and quantum gravity.

In addition, we further consider an EMR system to simulate the gravitational waves radiated by a single periodic orbit. Through the adiabatic approximation method and under the condition of fixing the hair parameter \(Q_m\) and parameter \(k\), we obtain the gravitational wave waveforms ($h_+$ and $h_-$) in two polarization directions.
The results show that the gravitational wave signal radiated by a single periodic orbit can accurately identify the zoom-whirl behavior of the orbit (see Fig. \ref{j}). Moreover, a larger hair parameter \(Q_m\) and a smaller parameter \(k\) will suppress the periodicity of the gravitational wave signal, increase the signal amplitude, and significantly change its phase (see Figs. \ref{k} and \ref{l}). This indicates that the hair parameters of the short hair BH have a significant impact on the spacetime structure near the event horizon and modulate the gravitational wave signal through quantum effects. Future gravitational wave detections can not only reveal these orbital characteristics but also further reveal the internal structure and orbital evolution of the short hair BH, providing a potential observational approach for verifying the no-hair theorem or the "no-short-hair" theorem.

In this paper, the short hair BH model we study is static and spherically symmetric, although its physical background is similar to the real-universe environment (that is, the case of considering anisotropic matter instead of vacuum). However, it is well-known that BHs in the real universe usually have spins. Therefore, although the static BH model may be a good approximation in the region far from the BH, when studying the particle motion close to the BH's event horizon, the introduction of spin effects may reveal more interesting physical phenomena. In particular, the frame-dragging effect caused by spin may enhance the quantum effect and significantly affect the gravitational wave signal or the orbital behavior of test particles, which may lead to novel observational results. Therefore, considering spin effects and reconstructing more accurate gravitational wave waveforms are of great significance for in-depth understanding of the physical properties of the short hair BH.
In future research, we plan to further explore the complex property of the short hair BH by analyzing the timelike geodesic dynamics in the rotating short hair BH and constructing precise gravitational wave waveforms (such as taking high-order multipole moments and non-adiabatic situations into consideration).

\section{acknowledgements}
Weacknowledge the anonymous referee for a construc tive report that has significantly improved this paper. This work was  supported by Guizhou Provincial Basic Research Program(Natural Science)(Grant No. QianKeHeJiChu-[2024]Young166),  the Special Natural Science Fund of Guizhou University (Grant No.X2022133), the National Natural Science Foundation of China (Grant No. 12365008) and the Guizhou Provincial Basic Research Program (Natural Science) (Grant No. QianKeHeJiChu-ZK[2024]YiBan027) .


\bibliography{ref}
\bibliographystyle{apsrev4-1}

\end{document}